\documentclass[journal=jacsat,manuscript=article]{achemso}
\SectionNumbersOn 

\usepackage[version=3]{mhchem} 
\usepackage{siunitx}           
\usepackage{amssymb}
\usepackage{multirow}          
\usepackage{graphicx}          
\usepackage{subfig}
\usepackage[T1]{fontenc}
\usepackage{hyperref}
\usepackage[format=hang]{caption}
\usepackage[numbers]{natbib}
\usepackage{xcolor}
\usepackage{soul}               

\usepackage{bm} 
\usepackage{physics} 

\newcommand* \aid{\hat{a}^\dagger_i}

\newcommand* \asd{\hat{a}^\dagger_s}
\newcommand* \vac{\ket{0}}
\newcommand* \ptpa{\mathcal{P}_{\text{TPA}}}

\setkeys{acs}{email=false}

\author{Aleksa Krsti\'c}
\affiliation[FSU-S]{Institute of Applied Physics, Abbe Center of Photonics, Friedrich Schiller University Jena, Albert-Einstein-Straße 15, 07745 Jena, Germany}

\author{Tobias Bernd G\"abler}
\affiliation[IOF]{Fraunhofer Institute for Applied Optics and Precision Engineering IOF, Albert-Einstein-Straße 7, 07745 Jena, Germany}
\alsoaffiliation[FSU-S]{Institute of Applied Physics, Abbe Center of Photonics, Friedrich Schiller University Jena, Albert-Einstein-Straße 15, 07745 Jena, Germany}

\author{Nitish Jain}
\affiliation[IOF]{Fraunhofer Institute for Applied Optics and Precision Engineering IOF, Albert-Einstein-Straße 7, 07745 Jena, Germany}

\author{Patrick Then}
\affiliation[FSU-E]{Institute of Applied Optics and Biophysics and Microverse Imaging Center, Friedrich Schiller University Jena, Philosophenweg 7,07743 Jena, Germany}
\alsoaffiliation[IPHT]{Leibniz Institute of Photonic Technology IPHT, Albert-Einstein-Straße 9, 07745 Jena, Germany}

\author{\\Valerio Flavio Gili}
\affiliation[IOF]{Fraunhofer Institute for Applied Optics and Precision Engineering IOF, Albert-Einstein-Straße 7, 07745 Jena, Germany}

\author{Sina Saravi}
\affiliation[FSU-S]{Institute of Applied Physics, Abbe Center of Photonics, Friedrich Schiller University Jena, Albert-Einstein-Straße 15, 07745 Jena, Germany}

\author{Frank Setzpfandt}
\affiliation[FSU-S]{Institute of Applied Physics, Abbe Center of Photonics, Friedrich Schiller University Jena, Albert-Einstein-Straße 15, 07745 Jena, Germany}
\alsoaffiliation[IOF]{Fraunhofer Institute for Applied Optics and Precision Engineering IOF, Albert-Einstein-Straße 7, 07745 Jena, Germany}

\author{\\Christian Eggeling}
\affiliation[FSU-E]{Institute of Applied Optics and Biophysics and Microverse Imaging Center, Friedrich Schiller University Jena, Philosophenweg 7,07743 Jena, Germany}
\alsoaffiliation[IPHT]{Leibniz Institute of Photonic Technology IPHT, Albert-Einstein-Straße 9, 07745 Jena, Germany}

\author{Markus Gr\"afe}
\affiliation[TUD]{Institute of Applied Physics, Technical University of Darmstadt, Schloßgartenstraße 7, 64289 Darmstadt}
\alsoaffiliation[IOF]{Fraunhofer Institute for Applied Optics and Precision Engineering IOF, Albert-Einstein-Straße 7, 07745 Jena, Germany}

\title[]
  {Enhancing entangled two-photon absorption of Nile Red via temperature-controlled SPDC}
 
\begin{document}

\let\thefootnote\relax\footnotetext{\hspace{-0.65cm}Authors to whom correspondence should be addressed:\\ Aleksa Krsti\'c, \href{mailto:aleksa.krstic@uni-jena.de}{aleksa.krstic@uni-jena.de}; Tobias Bernd G\"abler, \href{mailto:tobias.bernd.gaebler@iof.fraunhofer.de}{tobias.bernd.gaebler@iof.fraunhofer.de}}







\newpage
\begin{abstract}
  
  Entangled two-photon absorption can enable a linear scaling of fluorescence emission with the excitation power. In comparison to classical two-photon absorption with a quadratic scaling, this can allow fluorescence imaging or photolithography with high axial resolution at minimal exposure intensities. However, most experimental studies on two-photon absorption were not able to show an unambiguous proof of fluorescence emission driven by entangled photon pairs. On the other hand, existing theoretical models struggle to accurately predict the entangled two-photon absorption behavior of chemically complex dyes. In this paper, we introduce an approach to simulate entangled two-photon absorption in common fluorescence dyes considering their chemical properties. Our theoretical model allows a deeper understanding of experimental results and thus the occurrence of entangled two-photon absorption. In particular, we found a remarkable dependency of the absorption probability on the phase-matching temperature of the nonlinear material. Further, we compared results of our theoretical approach to experimental data for Nile Red.
  
\end{abstract}


\section{Introduction}

Spectroscopic investigations of chemical and biological samples via the absorption of entangled photon pairs can offer several advantages in comparison to methods based on classical single- and two-photon absorption \cite{Javanainen1990,dorfman2016nonlinear,schlawin2017theory,Schlawin2018,Raymer2021}. A hallmark of entangled two-photon absorption (eTPA) is a linear dependence between excitation light intensity and absorption rate \cite{Javanainen1990}, which results in enhanced excitation rate at low excitation fluxes. This offers an attractive prospect in the context of biomedical imaging and microscopy, as lower excitation fluxes help reduce detrimental effects, e.g., photobleaching \cite{demchenko2020photobleaching}, caused by high power light sources required by techniques based on classical two-photon absorption (cTPA). 
With the emergence of efficient sources of entangled photon pairs based on nonlinear optical processes such as spontaneous parametric down-conversion (SPDC) \cite{couteau2018SPDCgeneral} in recent years \cite{kim2021studySPDC,zhang2021SPDCreview}, working in the low-excitation-flux regime has become much more accessible. This has led to the development of novel eTPA-based techniques for use in biomedical applications, where molecular dyes excited by entangled photons are becoming a topic of great interest \cite{Villabona-Monsalve2017,Schlawin2018,eshun2018DFT,Gu2021,Martinez-Tapia2023,Landes2021,Corona-Aquino2022}.

Despite this interest, the extremely low efficiency of multi-photon absorption processes in general and the highly complex spectroscopic behavior of common fluorophores present a number of difficulties in predicting, verifying, and quantifying the results of many eTPA experiments \cite{Martinez-Tapia2023,Landes2021,Corona-Aquino2022}. This, in turn, has led to the necessity of developing new systematic and application-oriented approaches, both theoretical and experimental, for describing and quantifying eTPA.
While there have been many experimental works on eTPA in recent years, reporting a wide range of behaviors in different dyes \cite{Villabona-Monsalve2017,Parzuchowski2021,Villabona-Monsalve2022,Hickam2022, Gaebler2023}, current theoretical works often focus on either fundamental quantum-metrological properties of the process {\cite{Sanchez2021, Panahiyan2022, Landes2021, Landes2022,Saleh1998, Schlawin2024}}, or photon-pair state optimization only for model or less complex systems \cite{schlawin2017theory,oka2018tpa_Rb, Leon-Montiel2019,Mertenskoetter2021, Svozilik2018}. However, quantitative predictions of expected absorption rates or cross-sections for specific dye molecules, relying on data obtained using quantum-chemical methods, have been performed in the context of cTPA \cite{Kang2020, Gu2021, hornum2020solvatochromism,guo2021NileRedDFT}. Those results are not easily applicable to eTPA whose properties are also heavily dependent on the properties of the source of entangled photons \cite{Landes2021,Landes2022,Raymer2021}. Nevertheless, a first study by Varnavski et al. was published in 2023, which demonstrated the possibility of ab-initio calculations for quantitative predictions of experimental eTPA scenarios {\cite{Varnavski2023}}. Unfortunately, their approach does not allow simple and fast predictions because of the requirement of computational chemistry software packages.

\begin{figure}[htpb]
\centering
\subfloat[Chemical structure of Nile Red]{\includegraphics[width=0.35\textwidth]{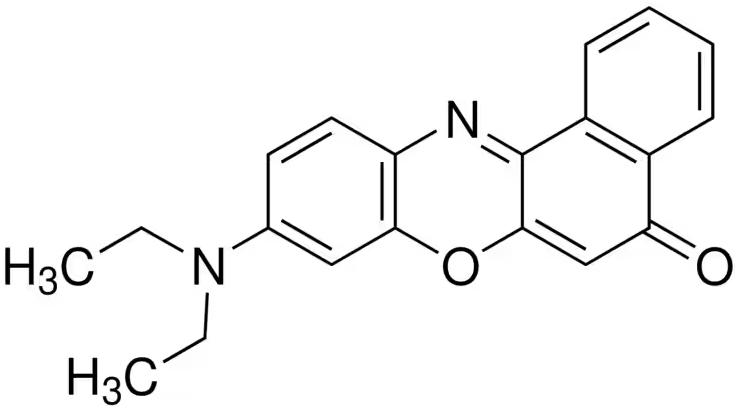}}\hfil
\subfloat[General model of the energy structure of a fluorophore]{\includegraphics[scale=1.05]{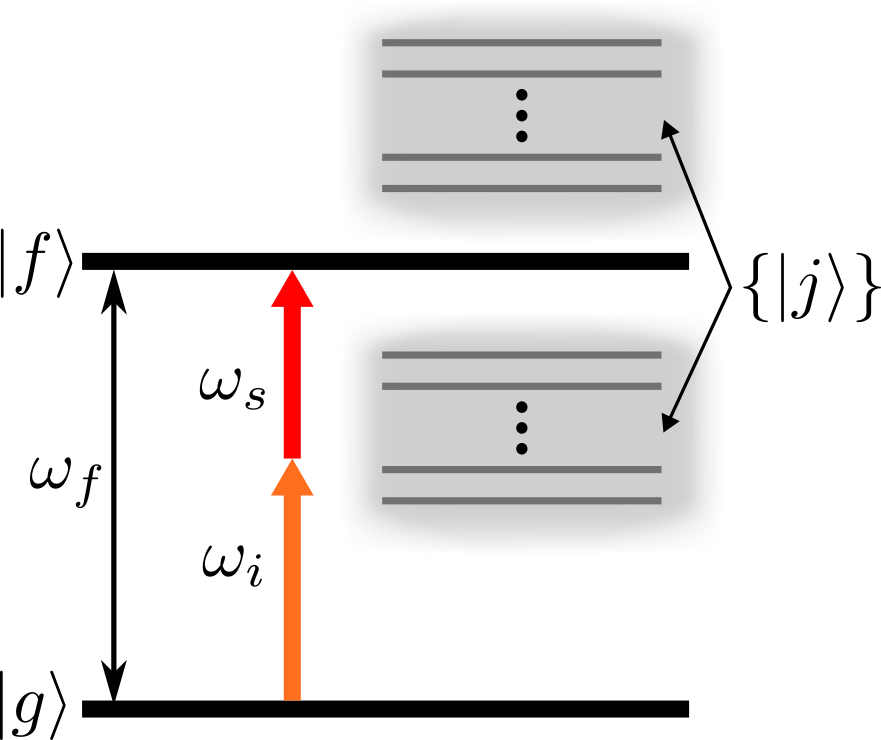}}

\caption{\textbf{(a)} Chemical structure of the investigated dye Nile Red. \textbf{(b)} Schematic representation of the model energy structure used in our theoretical modelling of eTPA. We assume fixed ground and excited states $\ket{g}$ and $\ket{f}$, respectively, but multiple states $\ket{j}$ which contribute to the absorption process
 as potential intermediate states.}\label{fig:figure_1}
\end{figure}

In this paper, we take steps to address this issue by presenting a systematic approach of estimating absorption probabilities of entangled photons for fluorescence dyes based on available molecular parameters and characteristics of the entangled photon source (Sec.~\ref{sec:Modelling eTPA}). To demonstrate the accuracy of our model and elaborate on its applicability, we compare results gained by our approach for the common fluorescence dye Nile Red, shown in Fig.~\ref{fig:figure_1}(a), with experimental data (Sec.~\ref{sec:Experiment}).

\section{Theoretical Modelling of eTPA} \label{sec:Modelling eTPA}

\subsection{Expression for the Probability}
We consider a model molecule with a level structure shown in Fig.~\ref{fig:figure_1}(b) interacting with a two-photon optical field, generated using a SPDC source. The molecule is initially in the ground state $\ket{g}$ and, by absorbing the photons from the field, is excited to the final state $\ket{f}$. We assume an incoming frequency-entangled two-photon state of the general form
\begin{align}\label{eq:fieldState}
\ket{\Psi}=\iint\dd{\omega_i}\dd{\omega_s}A(\omega_i,\omega_s)\aid\asd\vac,
\end{align}
where $\omega_{i,s}$ are the frequencies of the photons, which we refer as idler ($i$) and signal ($s$). $A(\omega_i,\omega_s)$ is the joint spectral amplitude (JSA) of the incoming two-photon state, describing the photons' spectral properties and correlations, and $\hat{a}_{i,s}^\dagger$ are the creation operators for the two photons. To calculate the probability $\ptpa$ of finding the molecule in the excited state after the interaction with the incident photons, we apply perturbation theory, which is a well-established approach for treating both cTPA and eTPA \cite{mollow1968two,Raymer2021,Landes2022,deMontiel2013roleOfCorrelations}. By assuming that the center frequencies of the incoming photons $\omega_{i0}$ and $\omega_{s0}$ are far from any intermediate resonances and their sum is nearly resonant with the two-photon transition frequency $\omega_f$, we can write the transition probability in the frequency domain, valid up to fourth order as \cite{deMontiel2013roleOfCorrelations,mollow1968two}:
\begin{align}\label{eq:probability}
    \ptpa\propto\int \dd{\omega_0}\left|\int \dd\omega_{i}\dd\omega_{s}\delta\left(\omega_{0}-\omega_{i}-\omega_{s}\right)L_{\omega_0}\left(\omega_{i},\omega_{s}\right)A\left(\omega_{i},\omega_{s}\right)\right|^{2},
\end{align}
where we introduced the \textit{two-photon response} of the molecular system:
\begin{align}\label{eq:2phResponse}
    L_{\omega_0}(\omega_i,\omega_s)=\sqrt{g\left(\omega_{0}\right)}\underset{j}{\sum}D_{j}\left(\frac{1}{\Delta_{j}\left(\omega_{i}\right)}+\frac{1}{\Delta_{j}\left(\omega_{s}\right)}\right).
\end{align}
The sum in Eq.~\ref{eq:2phResponse} iterates over all the possible intermediate states $\ket{j}$ which may be either real energy levels or virtual states \cite{Raymer2021,Landes2022,deMontiel2013roleOfCorrelations,Kang2020} with frequencies $\omega_j$, linewidths $\gamma_j$ and frequency mismatches $\Delta_j(\omega)=\omega_j-\omega+i\gamma_j$. $D_j=\mu_{gj}\mu_{jf}$ is the product of the transition dipole moments between the ground and intermediate states $\mu_{gj}$ and intermediate and excited states $\mu_{je}$, introduced for notational compactness. Furthermore, we assume that the final state is homogeneously broadened with a width $\Gamma_f$ \cite{mollow1968two,Landes2021}, which is included in Eq.~\ref{eq:2phResponse} as ${g(\omega)=\frac{1}{\pi}\frac{\Gamma_f^2}{(\omega_f-\omega)^2+\Gamma_f^2}}$.

The assumption that the incoming photons' center frequencies are far from any intermediate resonances also neglects single photon effects such as hot band absorption (HBA) {\cite{drobizhev2003HBA,Mikhaylov2022,Varnavski2023}}. It has recently been shown that HBA in particular can mimic the behavior of eTPA in some cases. An in-depth assessment of the intensity of HBA in a given system should be done to avoid it skewing the estimation of the overall eTPA cross-section {\cite{Mikhaylov2022, Varnavski2023}}. The contribution of HBA in our experimental setup is discussed and estimated in Sec.{~\ref{sec:HBA}}.

\subsection{Our qualitative approach}
In general, the sum in Eq.~\ref{eq:2phResponse} should involve all of the possible states that can serve as intermediates in the two-photon transition. However, when dealing with real molecular systems, exact values for the parameters $\omega_j$, $D_j$ and $\gamma_j$ are difficult to obtain without resorting to costly time-dependent density functional theory simulations \cite{eshun2018DFT,guo2021NileRedDFT,Kang2020,Gu2021,tuck2009abInitio}. In addition, molecular systems, like fluorophores, can exhibit different spectroscopic properties, depending on their environment \cite{hornum2020solvatochromism,martinez2016nileReview}, thus rendering the exact calculation of the two-photon response even more difficult. For example, the absorption maximum of Nile Red can vary by as much as $100$nm depending on the solvent used \cite{jose2006benzophenoxazine}. Despite these obstacles, we can apply a qualitative understanding of the two-photon response that relies on very few parameter values to infer the eTPA behavior of Nile Red and devise an optimization procedure for the excitation setup given in Sec.~\ref{sec:Experiment}.

\begin{figure}[htpb]
\centering
\subfloat[]{\includegraphics[scale=0.95]{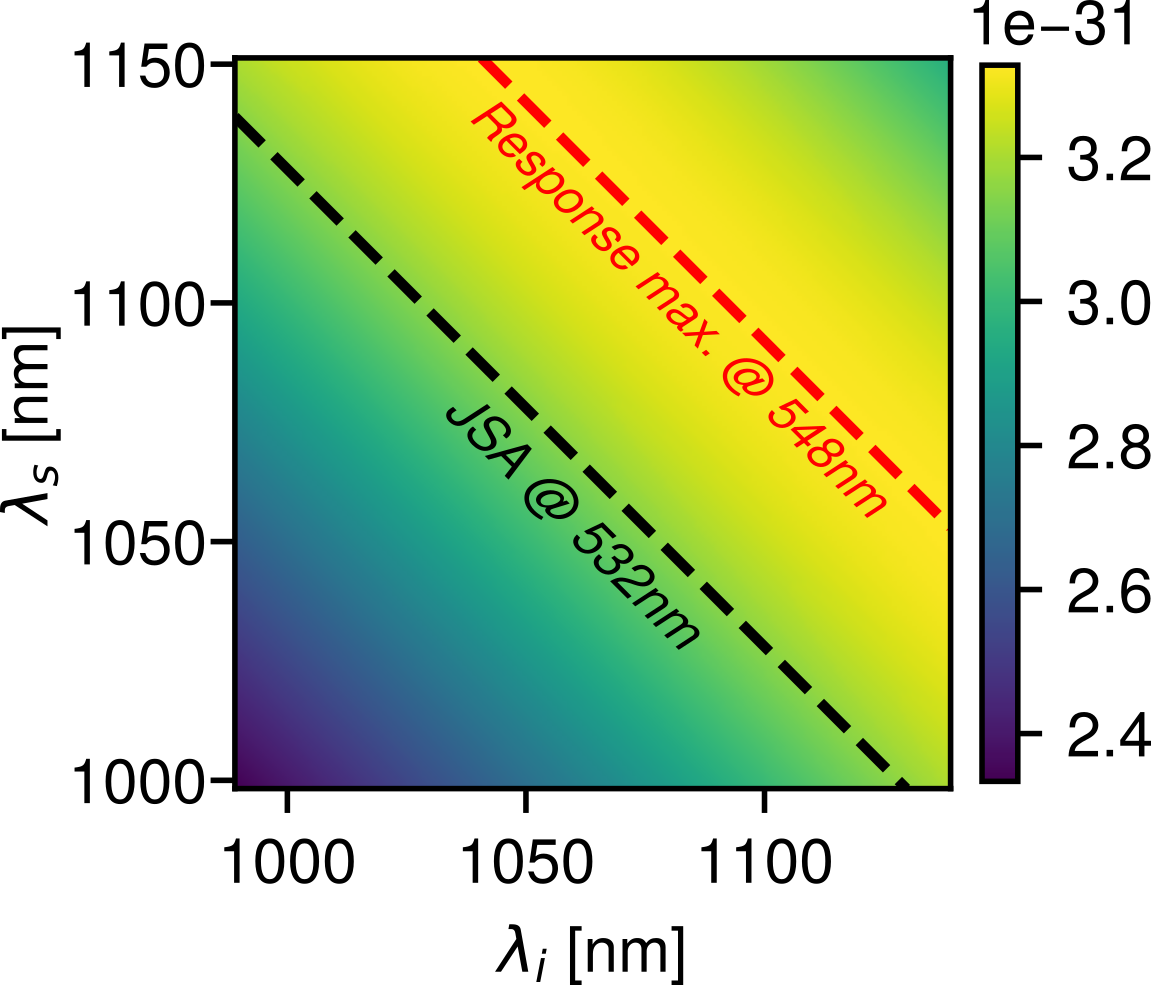}}\hfill
\subfloat[]{\includegraphics[scale=0.95]{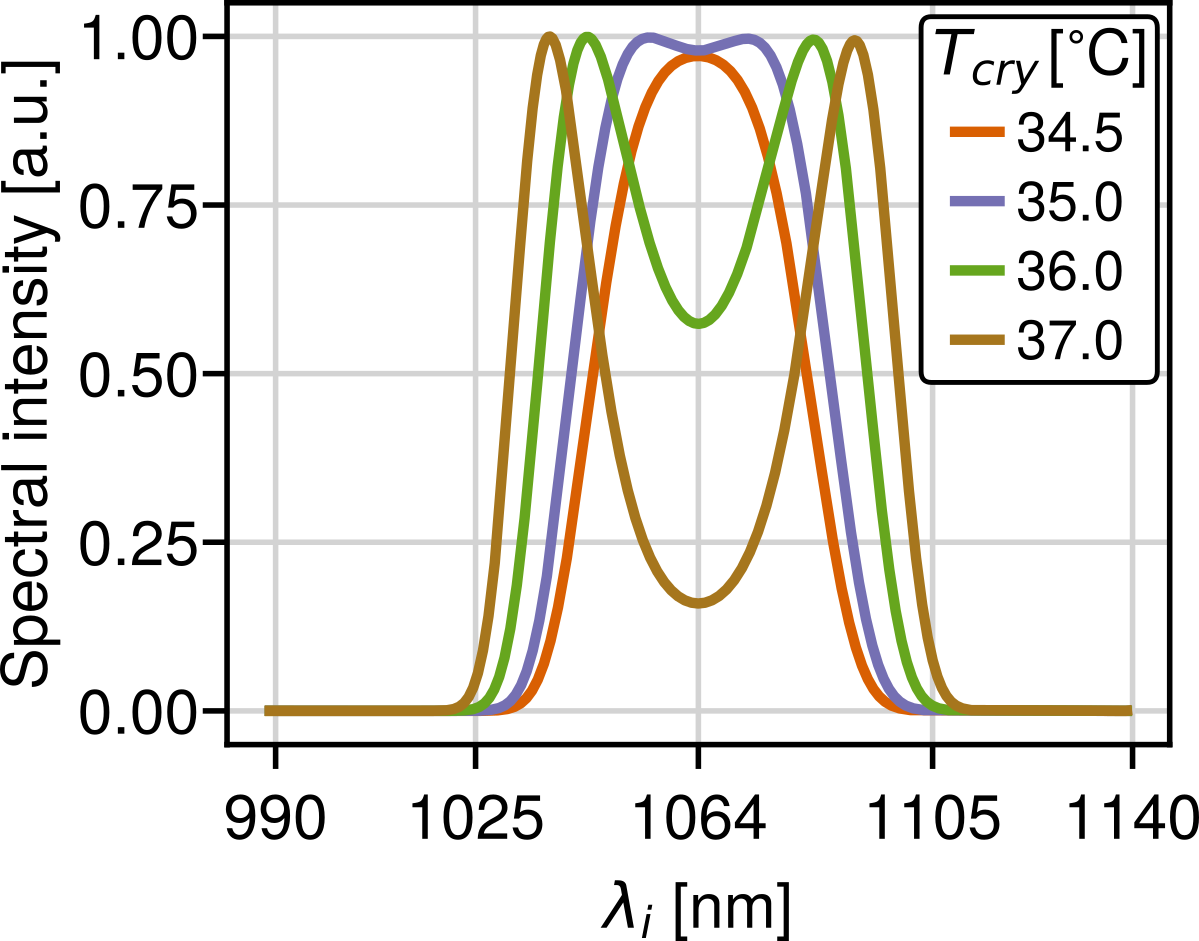}}\hfill
\subfloat[]{\includegraphics[scale=0.95]{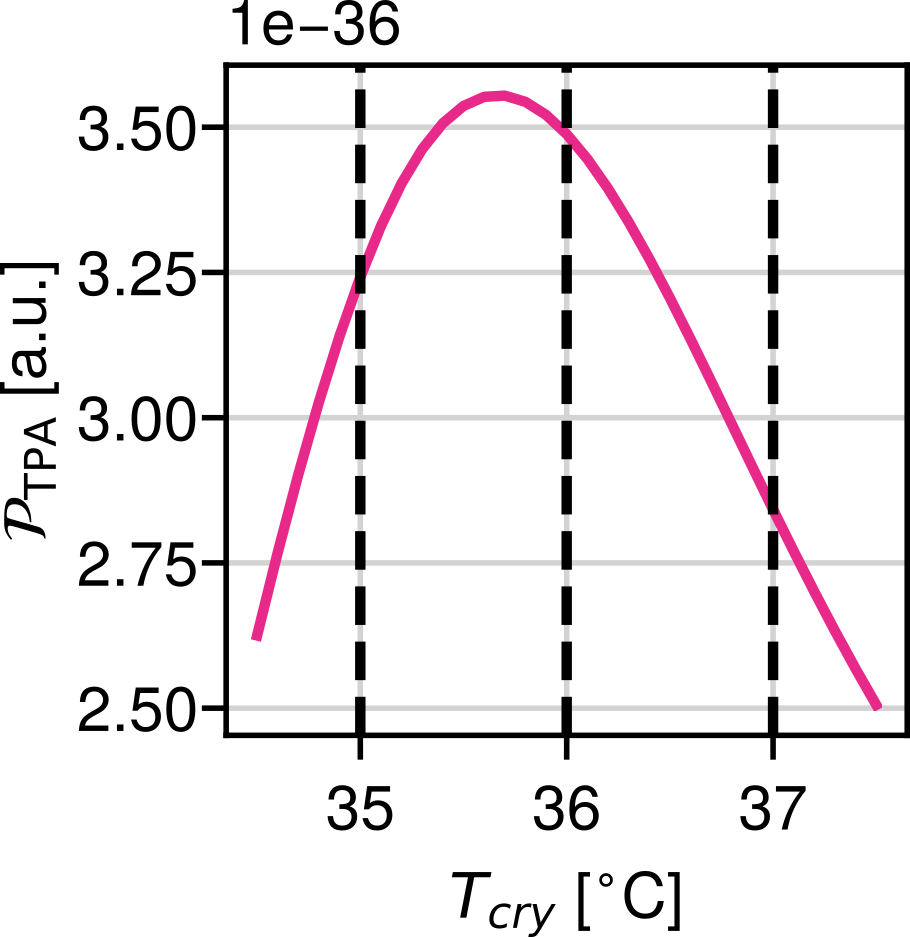}} 
\caption{\textbf{(a)} Modelled molecular response of Nile Red. The orange dashed line indicates the maximum of the calculated molecular response and the black dashed line indicates the location of the joint photon-pair spectrum for a pump center wavelength of $\SI{532}{\nano\metre}$. \textbf{(b)} The modelled single-photon spectrum of the SPDC source for different crystal temperatures, illustrating the spectral bandwidth generated at each temperature. \textbf{(c)} The probability of eTPA obtained from Eq.~\ref{eq:probability} and the modelled molecular response and JSA. The dashed black lines indicate the nonlinear crystal temperatures investigated in Sec.~\ref{sec:Experiment}.}\label{fig:figure_2}
\end{figure}

Spectroscopic and quantum-chemical investigations of Nile Red suggest the commonly accessed absorption line at $\sim\SI{540}{\nano\metre}$ to be the lowest-energy optically active transition \cite{tuck2009abInitio,dias1999theoreticalNileRed,dias2006semiEmpirical,martinez2016nileReview,selivanov2011theoretical,nileRedAbsorptionSpectrum}, hence, only levels above the first excited state participate as intermediates in the eTPA process. This has a significant impact on the spectral properties of the molecular response, which determines the optimal two-photon excitation spectrum \cite{schlawin2017theory,krstic2019towards}. In Fig.~\ref{fig:figure_2}(a), we show the absolute value of $L_{\omega_0}(\omega_i,\omega_s)$ for $\omega_0=\omega_i+\omega_s$ over a range of wavelengths relevant to our experimental investigation. While the energies and dipole moments of the first few excited states of Nile Red have been determined in multiple numerical studies\cite{tuck2009abInitio,dias1999theoreticalNileRed,dias2006semiEmpirical}, the state-widths and transition dipole moments are more difficult to obtain \cite{Kang2020}. With these limitations in mind, to obtain the response shown in Fig.~\ref{fig:figure_2}(a), we made the following assumptions: in addition to the target state at $\lambda_f=\SI{548}{\nano\metre}$, we included two higher excited states, at $\lambda_1=\SI{440}{\nano\metre}$ and $\lambda_2=\SI{325}{\nano\metre}$\cite{hornum2020solvatochromism,guo2021NileRedDFT}, since all states above those two are detuned by more than $\omega_f/2$ from the investigated transition and their influence can be safely neglected. $\Gamma_f$ was set to $2\pi\times\SI{50}{\tera\hertz}$ (corresponding to a bandwidth of $\SI{50}{\nano\metre}$), in accordance with theoretical and experimental investigations of Nile Red's TPA spectrum \cite{hornum2020solvatochromism,guo2021NileRedDFT}. The values of $\gamma_{1,2}$ were both fixed to $2\pi\times\SI{24}{\tera\hertz}$, corresponding to the commonly accepted empirical assumption of $\sim\SI{0.1}{\eV}$ for intermediate state widths in cases where the exact values are not known \cite{beerepoot2015linewidths,Kang2020}. Finally, the transition dipole moment products $D_j=\mu_{gj}\mu_{jf}$ were set to $D_1=0.086$ and $D_2=0.078$, given here as unitless quantities since these values were calculated from oscillator strengths (which are unitless) for their respective transitions given in other works on Nile Red \cite{tuck2009abInitio,selivanov2011theoretical}. In practice, their units are absorbed into the overall proportionality factor implied in Eq.~\ref{eq:probability}.

In our simulations, we observed that the exact numerical values of $\gamma_j$ and $D_j$ have a limited impact on the spectral behavior of $L_{\omega_0}(\omega_i,\omega_s)$, as a consequence of the intermediate states lying above the target state\cite{krstic2019towards}. In particular, we found that increasing/decreasing $\gamma_j$ by a factor of $2$ has a negligible effect on the spectral shape and amplitude of $L_{\omega_0}(\omega_i,\omega_s)$. This indicates that the empirical assumption of a fixed value for each $\gamma_j$, which also neglects any broadening effects\cite{beerepoot2015linewidths}, does not negatively impact our qualitative predictions. On the other hand, we found that the values of $D_j$ do affect $L_{\omega_0}(\omega_i,\omega_s)$, but only its amplitude and that increasing/decreasing each of them by as much as $2$ orders of magnitude does not affect its spectral shape at all. From these observations, we conclude that qualitative predictions of a fluorophore's eTPA response and its spectral behavior are possible, even in the absence of exact numerical data on transition dipole moments and intermediate state linewidths, as long as the intermediates all lie above the target state.

\subsection{Photon-pair JSA modelling}\label{sec:Overlaps}
The optimal spectrum of the entangled photon-pair for eTPA should maximize the overlap between $A(\omega_i,\omega_s)$ and $L_{\omega_0}(\omega_i,\omega_s)$, as evident in Eq.~\ref{eq:probability}. This is done by matching the center frequencies of the individual photons with the transition frequency $\omega_f=\omega_{i0}+\omega_{s0}$, as well as aligning the frequency correlations of the signal and idler photons with the overall shape of $L_{\omega_0}(\omega_i,\omega_s)$ in frequency space \cite{schlawin2017theory,krstic2019towards}. To demonstrate how this can be done in practice, we investigate the spectral properties of an entangled photon pair generated by a temperature-controlled SPDC source and calculate the spectral overlap of the aforementioned photon-pair JSA and the molecular response at different temperatures of the source.

The JSA of a photon pair emitted by a type-$0$ SPDC source with a bulk crystal as the nonlinear medium, such as the periodically poled Lithium Niobate crystal used in the experimental setup described in Sec.~\ref{sec:Experiment}, can be analytically written as $A(\omega_i, \omega_S)=\alpha_P(\omega_i+\omega_s)\cdot \phi(\omega_i,\omega_s)$, where $\alpha_P(\omega)$ is the spectrum of the pump pulse exciting the nonlinear medium, and $\phi(\omega_i,\omega_s)$ is the phase-matching spectrum \cite{couteau2018SPDCgeneral,kim2021studySPDC}. For the pump, we assume a Gaussian spectrum of the form $\alpha(\omega)=\frac{1}{\sqrt{2\pi\sigma_P^2}}\exp{-\frac{\left(\omega-\omega_{P0}\right)^2}{2\sigma_P^2}}$, where $\omega_{P0}=2\pi\times\SI{564}{\tera\hertz}$ (equivalent to a wavelength of $\lambda_P=\SI{532}{\nano\metre}$) is the center frequency of the pump pulse and $\sigma_P=2\pi\times\SI{1.7}{\giga\hertz}$ is the pump bandwidth. These parameters of the pump pulse were chosen to match the properties of the pump laser used in Sec.~\ref{sec:Experiment}.

The phase matching function in a bulk, periodically poled nonlinear crystal has the form $\phi(\omega_i,\omega_s)=\text{sinc}\left(\frac{\Delta k(\omega_i,\omega_s)l}{2}\right)$. Inside the phase-matching function, $l$ is the length of the crystal and $\Delta k(\omega_i,\omega_s)$ is the phase mismatch determined by the photon frequencies, the refractive index of the crystal $n(\omega)$, and periodic poling. We model the temperature-dependent refractive index of a periodically poled, MgO doped Lithium Niobate crystal with a poling period of \SI{6.93}{\micro\metre} using a Sellmeier equation \cite{gayer2008refIndex} and use it to calculate the output spectrum of a type-0 SPDC source where the pump, idler and signal photons are all polarized along the crystal's extraordinary axis. This then allows us to calculate the probability Eq.~\ref{eq:probability} as a function of source temperature $T_{\text{cry}}$. 

To investigate the spectral distribution of the two-photon state at the output of the SPDC source, we plot the single-photon spectra of our source at different crystal temperatures in Fig.~\ref{fig:figure_2}(b). We do this because the chosen bandwidth of the pump pulse results in a JSA which is too narrow to be adequately shown here and the single-photon spectrum already contains all of the information required for our purposes. For clarity, we also indicate the position of the two-photon spectrum in Fig.~\ref{fig:figure_2}(a) as a black dashed line, defined by $\lambda_i+\lambda_s=\lambda_P$. Also indicated in Fig.~\ref{fig:figure_2}(a) is the maximum of the two-photon response, shown as an orange dashed line. Although the center of the JSA and the maximum of the two-photon response do not match exactly for the pump parameters considered presently, the resulting $\ptpa$ is not reduced significantly (as compared to the case where the two photons are perfectly resonant with the two-photon transition), only by around $6\%$. This kind of mismatch is to be expected in practice, and, as will be discussed below, does not influence the general dependence of $\ptpa$ on the spectral shape of the JSA.\\

The single-photon spectrum $S(\omega_i)$ of a photon-pair state is calculated as \linebreak$S(\omega_i)\propto\abs{\int\dd{\omega_s}A(\omega_i,\omega_s)}^2$ and the spectra shown in Fig.~\ref{fig:figure_2}(b) are given in terms of wavelength. From the spectra we see that for $T_{\text{cry}}\leq\SI{34.5}{\celsius}$ the source emits degenerate photons around $\SI{1064}{\nano\metre}$ and, as the temperature is increased, the photons become nondegenerate and the overall bandwidth of the SPDC process increases. At even higher temperatures ($T_{\text{cry}}>\SI{37}{\celsius}$), the spectra of the signal and idler photons become completely distinct. 

\subsection{The spectral overlap}\label{sec:SpectralOverlaps}
In Fig.{~\ref{fig:figure_2}}(c), we show the dependence of $\ptpa$ on $T_{\text{cry}}$ and see that the temperature-dependent bandwidth in the range $\SI{34}{\celsius}<T_{\text{cry}}<\SI{37}{\celsius}$ is responsible for maximizing the spectral overlap (and thus $\ptpa$) of the two-photon JSA and the molecular response. We observe that the probability rises as the SPDC bandwidth increases and maximizes for $T_{\text{cry}}=\SI{35.7}{\celsius}$, but starts decreasing as the photons become sufficiently nondegenerate and the shape of the spectrum results in a reduced overlap with the molecular response.

We can characterize this behavior by analyzing the spectral properties of $L_{\omega_0}(\omega_i,\omega_s)$ and the JSA. As shown in Fig.{~\ref{fig:figure_2}}(a), $L_{\omega_0}$ appears as a broad line along the antidiagonal corresponding to the absorption center wavelength of {\SI{548}{\nano\metre}}. In frequency space, this antidiagonal is defined by $\omega_i+\omega_s=\omega_f$. The described behavior is due to the absence of any optically active states between the ground and excited state. In contrast, the two photon JSA (in our source configuration) is very narrow, but is also oriented along an antidiagonal, one defined by $\omega_i+\omega_s=\omega_{P0}$. Even though $\omega_{P0}$ and $\omega_f$ differ, the excited state of the molecule is wide enough so that the two-photon response along the line on which the JSA lies is still comparable to its maximum value and each pair of photon frequencies in the JSA still "sees" a significant absorption probability. As can be seen from Fig.{~\ref{fig:figure_2}}(a) (by replacing the wavelengths with their corresponding frequencies), the values of $L_{\omega_0}$ along each antidiagonal direction remain approximately constant within the relevant frequency ranges (where the JSA is nonzero), centered around the degenerate frequency $\omega_{P0}/2$. This, along with the fact that the width of the JSA in the diagonal direction is much narrower than the linewidth of the final state $\Gamma_f$, enables us to approximate $L_{\omega_0}$ with a constant value $L_{\omega_{P0}} (\omega_{P0}/2,\omega_{P0}/2)$ over the whole JSA. This results in the spectral overlap being approximately $\ptpa\propto\int \dd \omega_0 \abs{\int \dd\omega_i A(\omega_0-\omega_i,\omega_i)}^2$, whose behavior we can investigate in different temperature regimes.

For source temperatures below around $\SI{35}{\celsius}$, the photons are spectrally degenerate and, as the phase matching bandwidth $\sigma_{PM}$ is much larger than the bandwidth of the pump $\sigma_P$, we can approximate the JSA as a "double-Gaussian" {\cite{grice2001eliminating,Raymer2021}}. Thus, it becomes a product of two Gaussian distributions, one along the diagonal direction $\omega_i=\omega_s$ with a width equal to $\sigma_P$, and a second one along the antidiagonal direction $\omega_i+\omega_s=\omega_{P0}$, with a width $\sigma_{PM}$. One can then evaluate the simplified expression for $\ptpa$ analytically and obtain that the spectral overlap is proportional to $\sigma_{PM}$. Thus indicating that the probability of eTPA is enhanced by increasing the phase matching bandwidth in the case of degenerate photons. This is confirmed by our results shown in Fig.{~\ref{fig:figure_2}}(c) as $\sigma_{PM}$, as well as the spectral overlap, increase with temperature in this temperature range.\\
For temperatures above $\SI{36}{\celsius}$, the photons are spectrally non-degenerate and the JSA consists of two distinct peaks. However, due to the indistinguishability of the signal and idler photons produced in a type-0 SPDC process, the spectral correlations of the state are embedded in both peaks equivalently and it is sufficient to consider either of them as the "new" JSA {\cite{Lerch2013tuning}}. Thus, the double-Gaussian approximation can now be applied again, and used to define the $\sigma_P$ and $\sigma_{PM}$ bandwidths of the new JSA.
The phase-matching bandwidth $\sigma_{PM}$ decreases with temperature in this regime {\cite{steinlechner2014efficient}} and, conversely, the eTPA probability also decreases.\\
In the intermediate temperature range $\SI{35}{\celsius}<T_{cry}<\SI{36}{\celsius}$, the transition between the two regimes of the JSA discussed above takes place. The maximal value of the eTPA spectral overlap, which also occurs in this temperature range, can likely be associated with the maximal value of the single-peak bandwidth that is attained just before the non-degenerate behavior of the JSA becomes dominant. 

The analysis given here and the conclusions made about the dependence of eTPA on the bandwidth of the photons' JSA is in agreement with results shown in other works {\cite{Raymer2021}} and confirms that a spectrally degenerate two-photon state with a high degree of spectral correlation is required for enhancing the efficiency of eTPA in molecular dyes with no active intermediate states.

\subsection{Validity of qualitative assumptions and estimation of HBA effects}\label{sec:HBA}
While the results presented here were obtained for Nile Red, our model and its methodology can be easily adapted for configurations involving other fluorophores and/or photon pair sources. Moreover, our simulations also showed the obtained dependence of $\ptpa$ on $T_{\text{cry}}$ to be quite robust to variations in the model parameters, namely $\gamma_j$ and $D_j$, whose exact values are not readily available in the literature. In particular, the functional dependence and optimal value of $T_{\text{cry}}$ remained the same, regardless of the parameter values, and only the overall amplitude varied, which does not influence the optimization procedure. This behavior, as well as well as the general dependence of $\ptpa$ on crystal temperature remains the same regardless of the detuning between the JSA spectrum and the maximum of the two-photon response, resulting from using a non-resonant pump. The detuning results in an overall decrease of $\ptpa$, as the amplitude of $L_{\omega_0}(\omega_i, \omega_s)$ is reduced for the detuned values of $\omega_{i,s}$, but, due to the large bandwidth of the final state, this decrease amounts to only a few $\%$ in the present case.

\begin{figure*}
    \centering
    \includegraphics{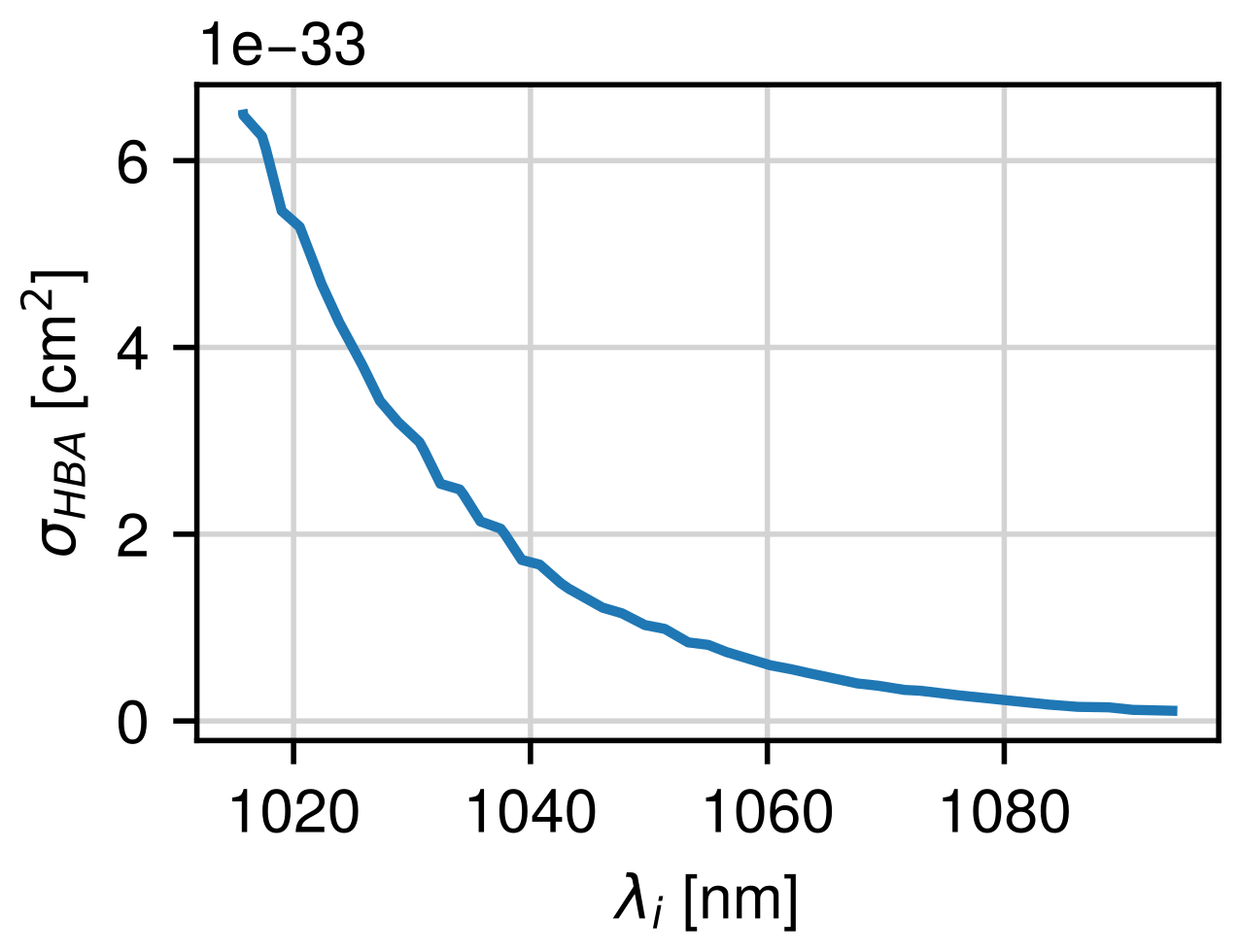}
    \caption{Estimated hot-band absorption cross-section $\sigma_{HBA}$ depending on the center wavelength $\lambda_i$ of the idler photon based on Eq.{~\ref{eq:HBA}}.}
    \label{fig:HBA}
\end{figure*}

As indicated previously, in order to estimate the overall contribution of HBA in our setup, we followed the methodology for estimating the HBA cross-section outlined in Ref.{~\cite{Mikhaylov2022,Varnavski2023}}. The cross-section $\sigma_{HBA}\left(\omega\right)$ is calculated as 
\begin{align}\label{eq:HBA}
    \sigma_{HBA}\left(\omega\right)&=\sigma_{0-0}\ e^{-\frac{\hbar}{kT}\left(\omega_{max}-\omega\right)}FC(\omega),
\end{align}
where $\omega_{max}$ is the frequency corresponding to the purely electronic $0-0$ transition, $\sigma_{0-0}$ is the single-photon absorption cross-section at $\omega_{max}$, $FC(\omega)$ is a normalized Franck-Condon factor, $T$ is the sample temperature, $\hbar$ is the reduced Planck constant and $k$ is Boltzmann's constant. The values for $\omega_{max}$, $\sigma_{0-0}$ and $FC(\omega)$ were determined using spectroscopic data {\cite{nileRedAbsorptionSpectrum}}, and the sample temperature was assumed to be {\SI{290}{\kelvin}}. Using the above expression, we were able to calculate an estimate for the HBA cross-section over the wavelength range of the incoming photons. Our calculations indicate that $\sigma_{HBA}$ in our setup is restricted to values smaller than \SI{6e-33}{\centi\metre^2} (see Fig.{~\ref{fig:HBA}}). When other experimental parameters and the measured photon flux of the source in our setup are taken into account (all of which are outlined in Sec.{~\ref{sec:Experiment}}), this cross-section results in an average number of photons lost due to HBA that is several orders of magnitude smaller than the absorption rates obtained in our measurements.

For the reasons given in this section, our qualitative approach to modelling the eTPA process represents a practical tool to estimate the success of eTPA experiments in advance. In addition, our model also provides a better understanding of the properties of the photon pair source relevant for eTPA applications.

\section{Comparison with Experimental Data}
\label{sec:Experiment}

\subsection{Experimental Setup}
\label{sec:Setup}

\begin{figure*}
    \centering
    \includegraphics{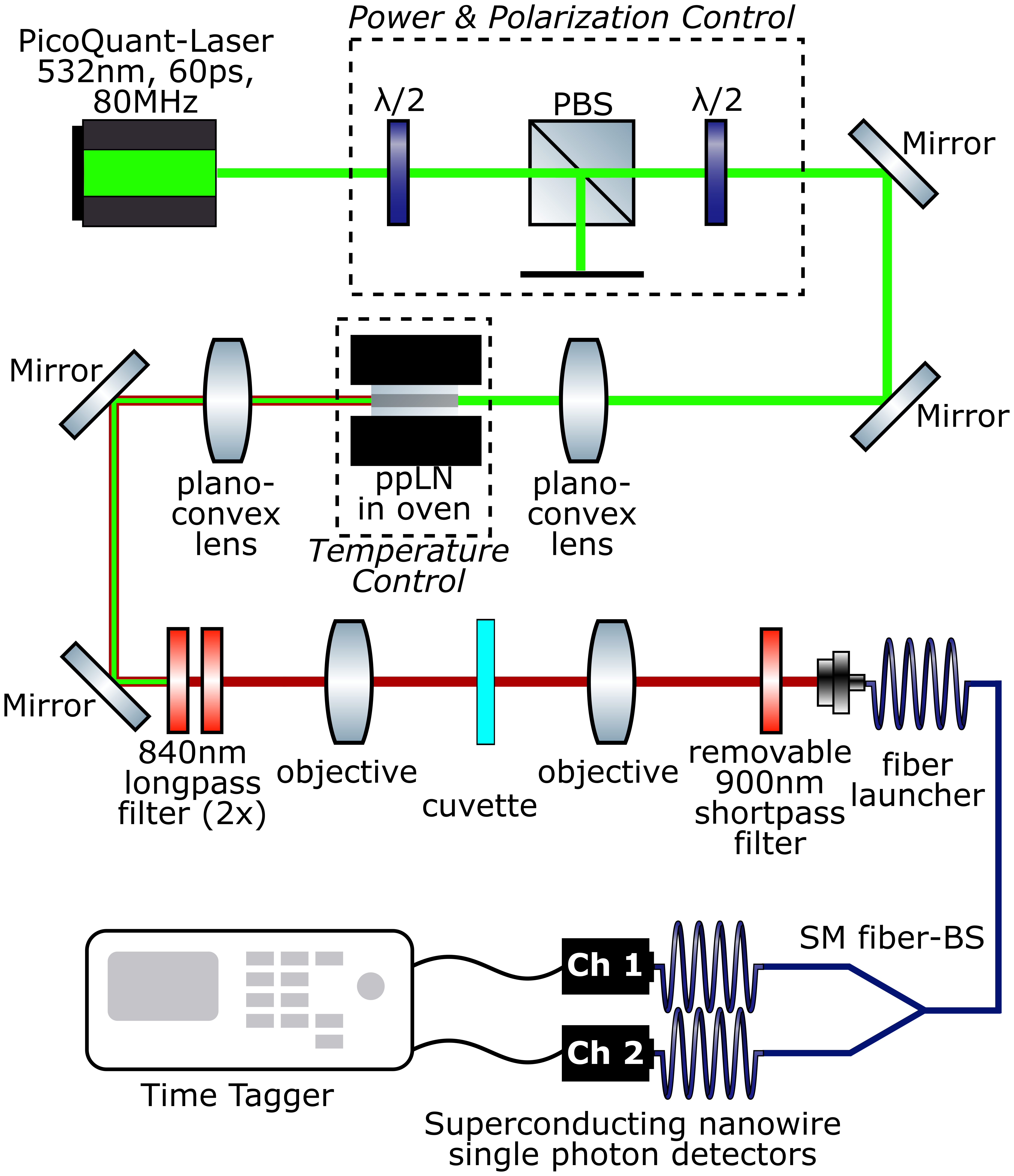}
    \caption{Experimental setup consisting of entangled photon pair source and a self-made transmission microscope}
    \label{fig:Setup}
\end{figure*}

To test the conclusions described in Sec.~\ref{sec:Modelling eTPA}, we measured the entangled two-photon absorption behavior of Nile Red and compared it with the behavior predicted by our theoretical analysis. For this purpose, we used the setup illustrated in Fig.~\ref{fig:Setup}. The entangled photon source is driven by a pulsed laser (PicoQuant VisUV-532) with a wavelength of \SI{532}{\nano\metre}, a linewidth of less than \SI{0.26}{\nano\metre}, a beam diameter of \SI{2.1}{\milli\metre}, a maximum average power of \SI{300}{\milli\watt}, a pulse length of \SI{60}{\pico\second} and a repetition rate of \SI{80}{\mega\hertz}. Power and polarization of this pump beam are controlled by two rotatable half-wave plates (Newport 10RP02-16) and a polarization beam splitter (Thorlabs PBS25-532-HP). The pump beam is guided by two mirrors (Thorlabs PF10-03-P01) and focused by a plano-convex lens (Thorlabs LA1608-A-ML) with a spot size of approximately \SI{27}{\micro\metre} into the non-linear medium where the entangled photon pairs are generated. As the nonlinear medium, we used a periodically poled, MgO doped Lithium Niobate crystal (Covesion MSHG1064-1.0) with a length of $l=\SI{20}{\milli\metre}$ and a poling period of \SI{6.93}{\micro\metre} inside an oven (Covesion PV20). Inside the crystal, entangled photon pairs with a center wavelength of \SI{1064}{\nano\metre} were generated via the process of SPDC. After passing through the crystal, the pump photons were removed by two filters (Semrock BLP01-830R-25) and the beam of entangled photons was collimated by a lens (Thorlabs LA1608-C) and guided by two mirrors (Thorlabs BB111-E03).

The two photon absorption and fluorescence tests were performed by a transmission microscope setup consisting of two opposed objectives (Olympus UPLFLN 10X/0.30). All investigated samples and solvents were placed between the objectives in cuvettes with a optical path length of $L=\SI{2.00}{\milli\metre}$ (Thorlabs CV2Q07AE). A removable shortpass filter (Thorlabs FESH0900) after the second objective allowed the separate consideration of fluorescence and absorption. The collection of transmitted photon pairs and emitted fluorescence photons was performed by a self-made fiber launch solution. For absorption measurements, the transmitted photons were coupled into a fiber beam splitter (Thorlabs TW1064R5F1B), which split and transferred them to two superconducting nanowire single-photon detectors (SNSPDs) (Single Quantum Eos 1064 CS). The respective single and coincidence rates detected by the SNSPDs were recorded via a PicoQuant MultiHarp 150.

\begin{figure}[htpb]
    \centering
    \subfloat[]{\includegraphics[width=0.4\textwidth]{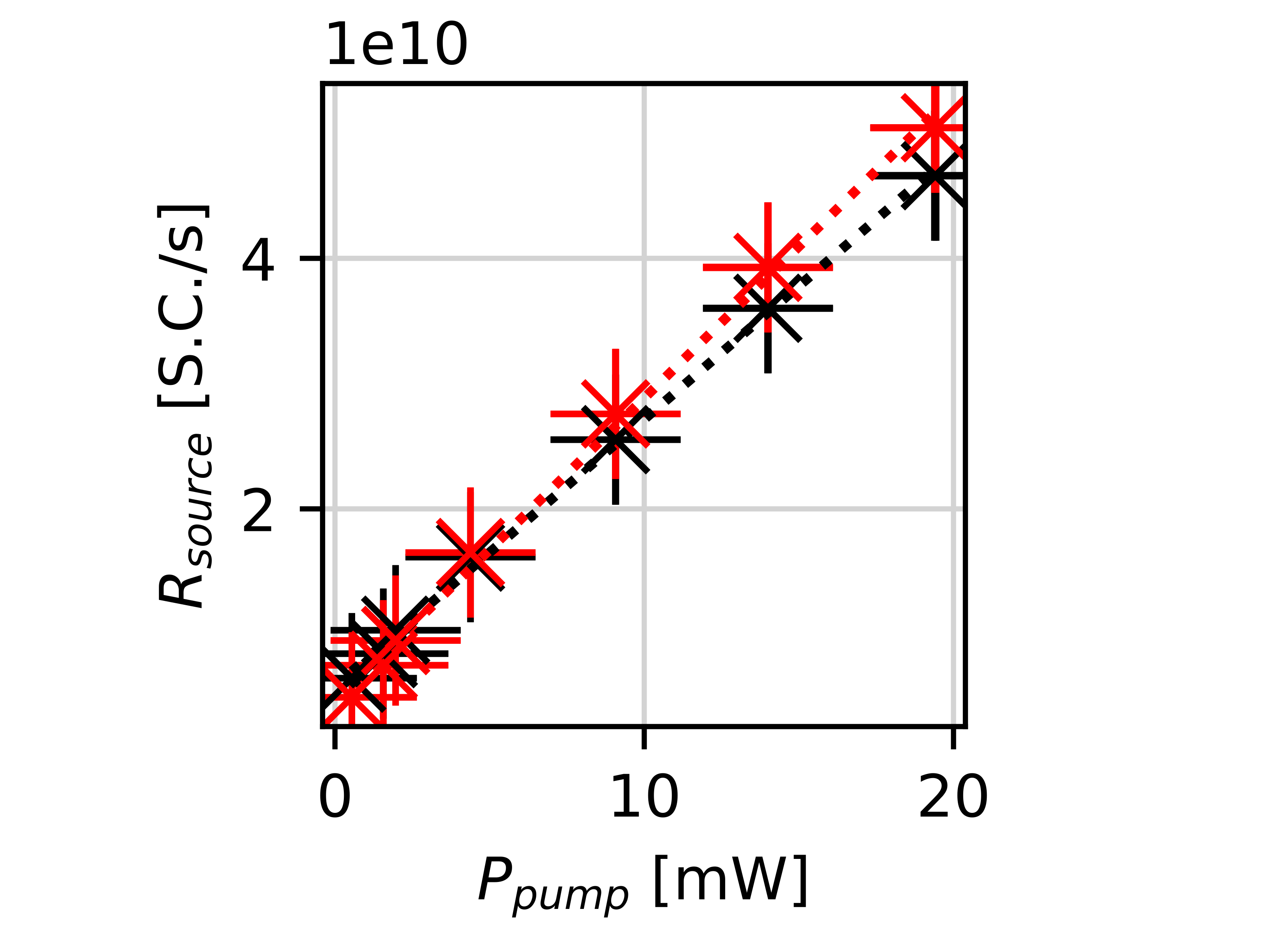}}
    \subfloat[]{\includegraphics[width=0.4\textwidth]{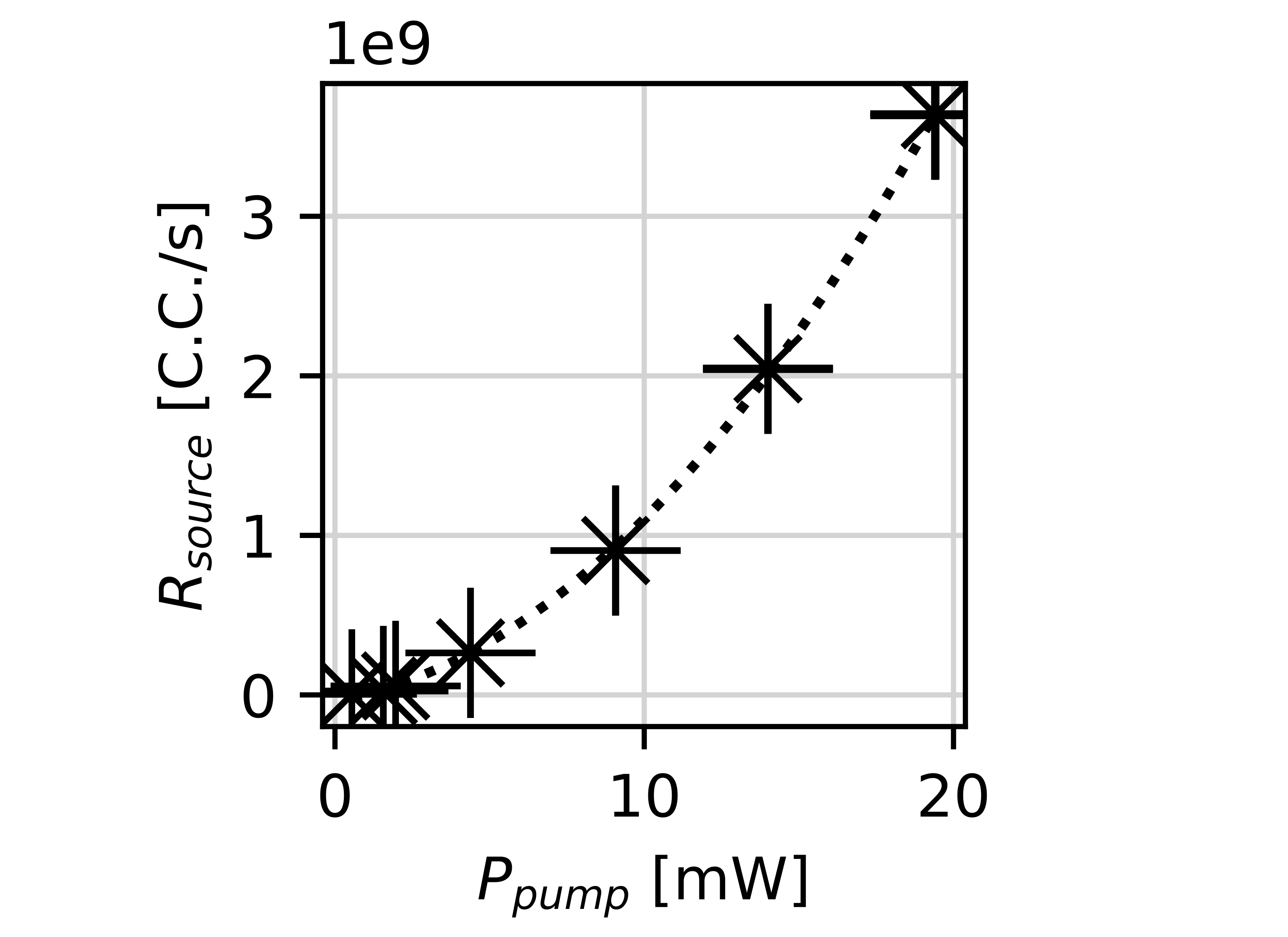}}
    \caption{\textbf{(a)} Single counts (black: channel 1; red: channel 2) and \textbf{(b)} coincidence counts generated by the described photon pair source for a crystal temperature of {\SI{35}{\celsius}} depending on the pump power (measured before the crystal).}
    \label{fig:SourceCounts}
\end{figure}

Fig.{~\ref{fig:SourceCounts}} shows the single and coincidence counts generated by this photon pair source. All values are corrected by dark counts and accidentals. For these measurements, the shortpass filter and both microscope objectives were removed from the setup, so that a collinear beam of photon pairs was coupled into the fiber beam splitter without any further interaction with disturbing optical elements. Additionally, the pump laser was attenuated with a neutral density filter (OD 0.3) to allow the recording of more data points before the used detectors saturated. The average pump power was measured by a powermeter (Thorlabs PM100D and Thorlabs S121C) over a period of one minute per data point. These recorded photon fluxes correspond to a maximum power of around {\SI{19}{\nano\watt}} at a wavelength of {\SI{1064}{\nano\metre}} for a pump power of approximately {\SI{20}{\milli\watt}}.

\subsection{Fluorescence Measurements}
\label{sec:Fluorescence}

To estimate the chances of success for detection of eTPA-induced fluorescence, we determined the lower limit to detect significant cTPA-induced fluorescence signals in an optical optimized and commercially available two-photon microscope. For this purpose, a Leica Stellaris 8 was used to investigate samples of dissolved Nile Red (Sigma-Aldrich 72485) in ethanol (Sigma-Aldrich 51976-500ML-F) illuminated by a laser (Spectra Physics InSight X3) with a center wavelength of \SI{1064}{\nano\metre}. Since the Leica Stellaris microscope used a combined detector system consisting of photomultiplier and avalanche diodes (Leica HyD NDD), we expected dark count per second in the lower two-digit region.

\begin{figure}
    \centering
    \includegraphics{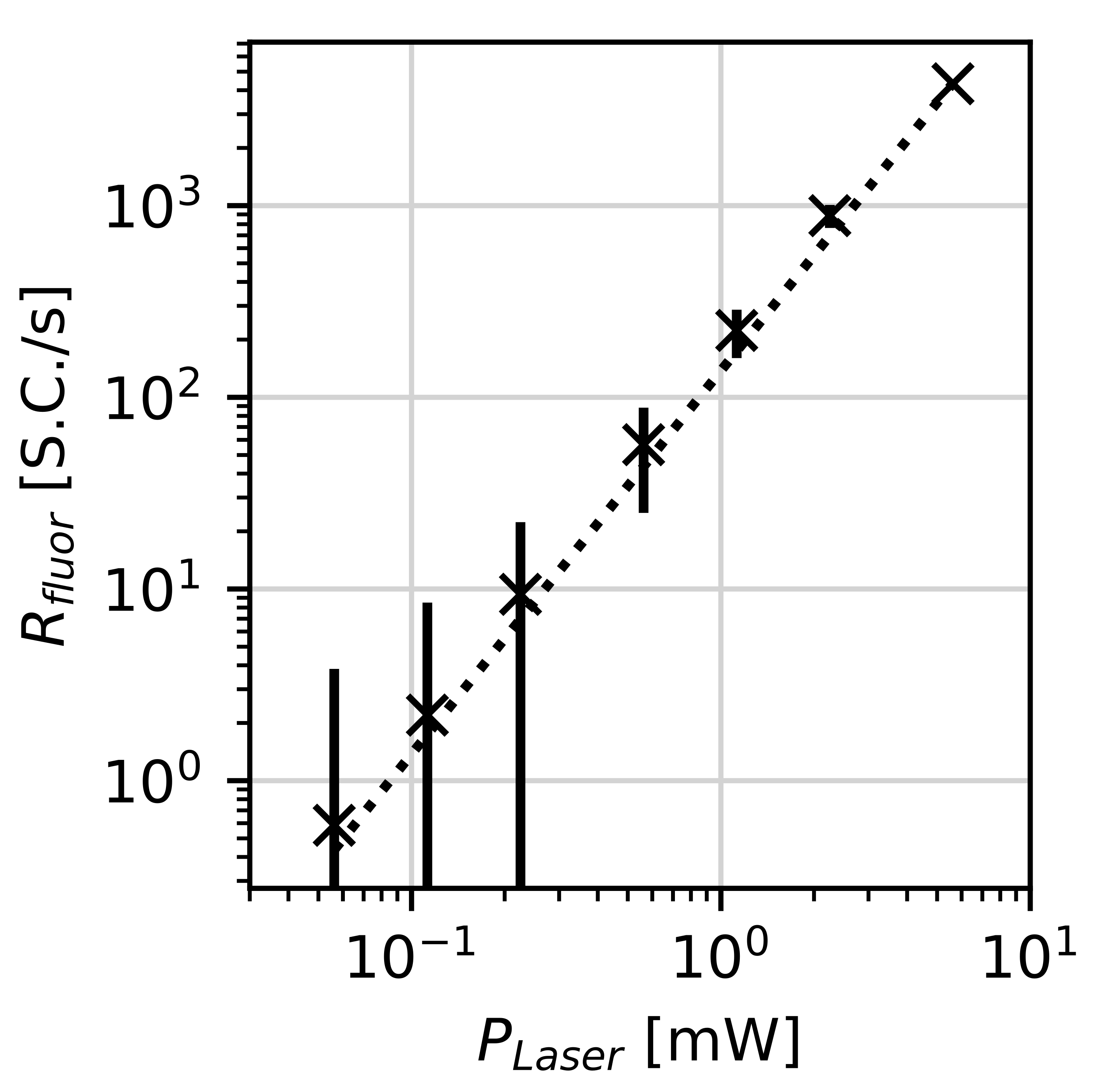}
    \caption{{cTPA-induced fluorescence rate $R_{fluor}$ of Nile Red, measured in single counts per second  (S.C./s), over the excitation} two-photon power $P_{Laser}$. The dotted line indicates the quadratic slope of the data as expected for cTPA.}
    \label{fig:Fluorescence}
\end{figure}

Fig.~\ref{fig:Fluorescence} shows the detected counts of fluorescence photons $R_{fluor}$ from a Nile Red sample with a concentration of $c=\SI{0.5}{\milli\mole\per\litre}$ excited with drastically reduced laser power $P_{Laser}$. A significant fluorescence signal was detected for an excitation power down to approximately \SI{1}{\milli\watt}. Further below, the signal-to-noise ratio became too low to get an unambiguous statement on whether the detected fluorescence followed cTPA, e.g., whether the power dependence clearly followed a quadratic dependence on the applied laser power.

Unfortunately, even if an enhancement of fluorescence excitation by eTPA will be assumed, this result already showed that the detection of eTPA-induced fluorescence becomes difficult in consideration of the several magnitudes lower light intensities generated by SPDC-based photon pair sources. This assumption was finally highlighted by our experiments with the setup described in Sec.~\ref{sec:Setup}. We were still not able to measure clearly identifiable eTPA-induced fluorescence using different detectors, for example a Si-based single photon avalanche detector (Excelitas SPCM-800-42-FC), the mentioned SNSPDs or a highly sensitive spectrometer (Ocean Insight QE Pro). This circumstance confirmed the problematic of showing a reliable experimental evidence of eTPA as mentioned in a large number of studies\cite{Parzuchowski2021, Corona-Aquino2022,Mikhaylov2022,Hickam2022,Gaebler2023,landes2024}. Because of the lack of fluorescence photons, we concentrated on absorption measurements in the following.

\subsection{Absorption Measurements}
\label{sec:Absorption}

To test the model described in Sec.~\ref{sec:Modelling eTPA} without the detection of eTPA-induced fluorescence, we now focused on the determination of the absorption cross-section $\sigma_{e}$ of entangled photons in Nile Red. For this purpose, we measured the transmission rates of the photon pairs for various pump power through the pure solvent $R_{solv}$ on the one hand and through the dissolved sample $R_{samp}$ on the other hand with the setup described in Sec.~\ref{sec:Setup}. All recorded data of $R_{solv}$ and $R_{samp}$ were corrected subsequently to account for detection and coupling efficiencies according to device specifications and measured dark and accidental counts. The subtraction $R_{solv}-R_{samp}$ for a given value of pump power finally correspond to the absorption pair rate $R_{abs}$.

\begin{figure}
    \centering
    \subfloat[]{\includegraphics[width=0.33\textwidth]{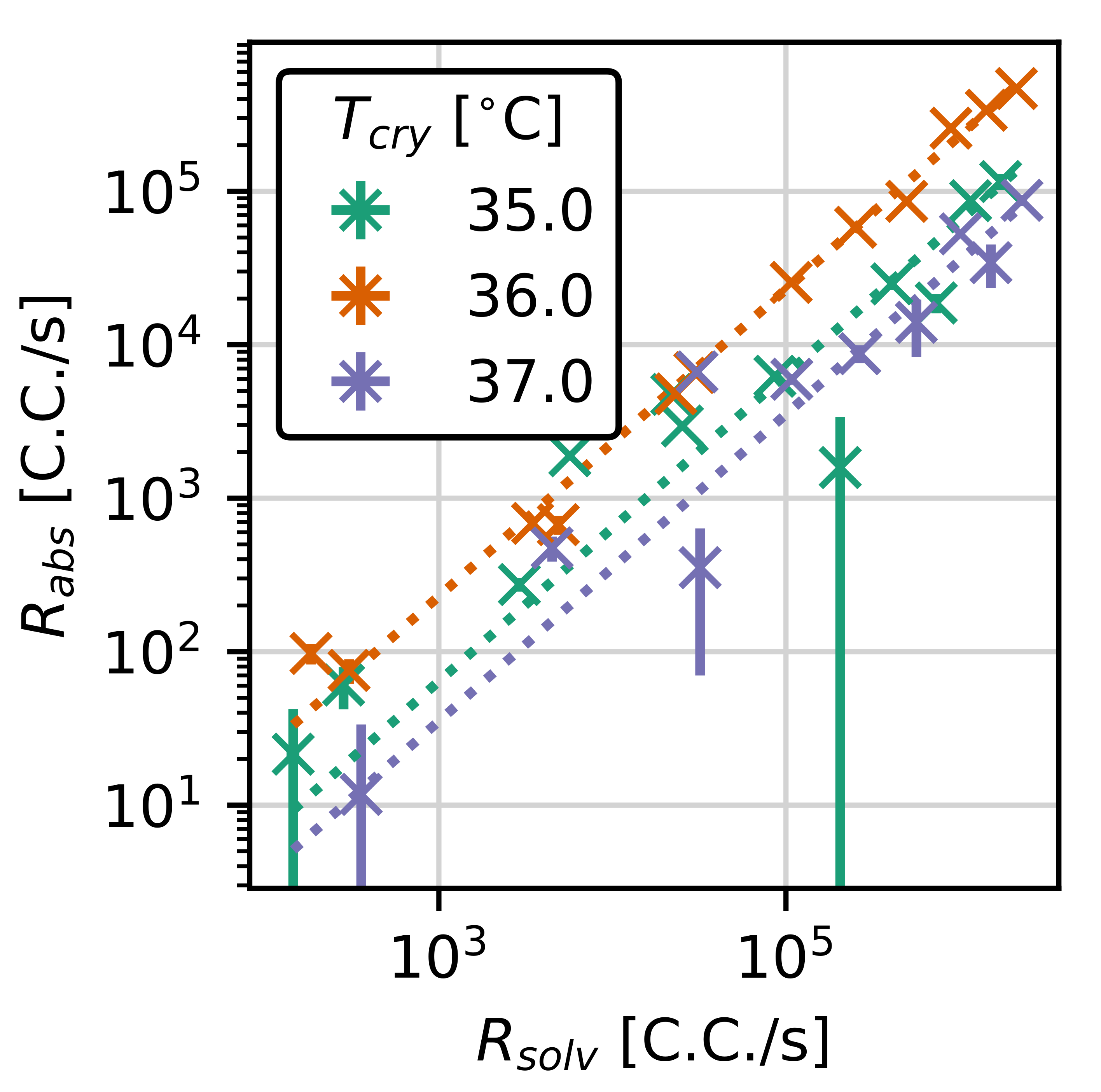}}
    \subfloat[]{\includegraphics[width=0.33\textwidth]{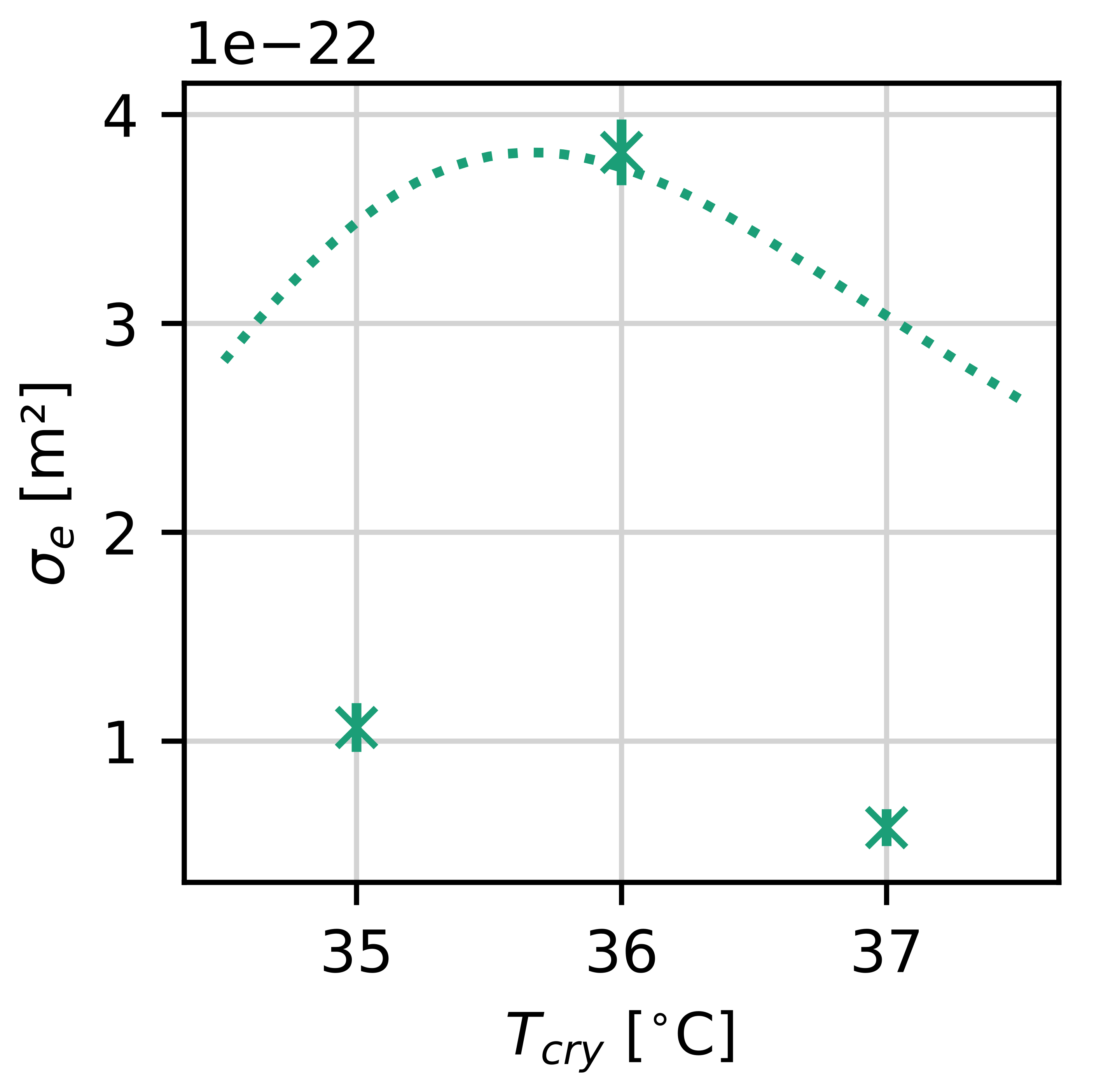}}
    \caption{\textbf{(a)} Absorption rates $R_{abs}$, measured in coincidence counts per second (C.C./s), over the according transmission rates $R_{solv}$ for different crystal temperatures $T_{cry}$. The dotted lines represent the slopes used for the calculation of entangled absorption cross-section $\sigma_{e}$. \textbf{(b)} $\sigma_{e}$ for the investigated values of $T_{cry}$. The dotted line represents the theoretical prediction shown in Fig.{~\ref{fig:figure_2}}(c) rescaled to the highest value of $\sigma_{e}$.}
    \label{fig:Absorption Rates}
\end{figure}

In Fig.~\ref{fig:Absorption Rates}(a), we show the dependence of $R_{abs}$ on the corresponding $R_{solv}$ for a Nile Red concentration of $c=\SI{0.5}{\milli\mole\per\litre}$ obtained for various temperatures of the nonlinear crystal $T_{cry}$. This dependence was then used to calculate the eTPA cross-section $\sigma_{e}$ as an experimental measure of the absorption probabilities $\ptpa$, defined as\cite{Corona-Aquino2022}
\begin{align}
    \sigma_{e}=\frac{1}{c\cdot L\cdot N_{A}}\cdot\frac{dR_{abs}}{dR_{solv}}.
    \label{eq:absorption cross-section}
\end{align}
Here, $c$ is the sample concentration, $L$ is the optical path length through the sample and $N_A$ is the Avogadro constant.  Finally, the slope $dR_{abs}/dR_{solv}$ is the main factor determining the eTPA cross-section. The values for $dR_{abs}/dR_{solv}$ obtained from fitting of the dependencies shown in Fig.~\ref{fig:Absorption Rates}(a) and $\sigma_{e}$ calculated using Eq.~\ref{eq:absorption cross-section} are given in Tab.~\ref{tab:Absorption cross-sections}.

\begin{table}
    \begin{tabular}[h]{c|c|c|c}
        $T_{cry}$ & \SI{35}{\celsius} & \SI{36}{\celsius} & \SI{37}{\celsius}\\ \hline
        $dR_{abs}/ dR_{solv}$ [1] & $0.064\pm0.005$ & $0.230\pm0.007$ & $0.035\pm0.003$ \\
        $\sigma_{e}$ [\SI{e-22}{\centi\metre^2}]& $1.1\pm0.2$ & $3.8\pm0.2$ & $0.59\pm0.09$\\ 
    \end{tabular}
    \caption{Calculated absorption cross-section $\sigma_{e}$ using Eq.~\ref{eq:absorption cross-section} based on the slopes illustrated in Fig.~\ref{fig:Absorption Rates}(a).}
    \label{tab:Absorption cross-sections}
\end{table}

The cross-section, as a function of the crystal temperature, is shown in Fig.{~\ref{fig:Absorption Rates}}(b), along with a rescaled plot of the theoretical dependence of $\ptpa$. The theoretical curve was obtained by fitting the points in Fig.{~\ref{fig:figure_2}}(c) with a fifth-order polynomial to obtain an approximate, closed-form, functional dependence of the eTPA probability to better compare it to experimental results. It is evident that the experimental data deviates from the theoretical predictions. While the measured cross-section does maximize at  approximately {$\SI{36}{\celsius}$}, the values drop away from the maximum much faster than the theoretical data predicts. This, in addition to the limited number of measured data points, prevents a clear final evaluation of the theoretical model.

In light of the work performed in Ref.~\cite{Martinez-Tapia2023}, where the ability of different experimental setup configurations to distinguish single- from two-photon effects in eTPA experiments was investigated, we note that our chosen configuration is not very well optimized for this distinction. In our experiment, the transmitted photon pairs enter the beam splitter at the same port, which does not enable a proper separation of single- and two-photon effects. A better option for this distinction offers N00N-state configurations. But due to the usage of type-0 SPDC in the near-degenerate regime, the necessary split of photon pairs by polarization or wavelength was not feasible so that a N00N-state configuration of the setup could not be used. Because of this fact, it cannot be ruled out that parasitic single-photon effects, as for example described in Ref. ~\cite{Hickam2022}, influenced the shown measurements.

A possible approach to investigate the occurrence of single- and two-photon effects is described by Ref.{~\cite{Corona-Aquino2022}}. It introduced the biphoton ratio $\Gamma$.
\begin{align} 
    \Gamma &= 1-\frac{r^{(1)}_\mathrm{samp}r^{(2)}_\mathrm{samp}/R_\mathrm{samp}}{r^{(1)}_\mathrm{solv}r^{(2)}_\mathrm{solv}/R_\mathrm{solv}} \label{eq:Biphoton ratio}
\end{align}
This quantity consists of the transmitted photon count rates $r$ (for singles) and $R$ (for coincidences) through the sample (samp) and solvent (solv). The upper indices of $r$ represents the channel of photon detection. Because of the inclusion of single as a well as coincidence count rates in $\Gamma$, it indicates different behaviors of single photons and photon pairs if two-photon effects during the absorption process exists. In doing so, a value close to zero implies the absence of two-photon effects, whereas values deviating from zero may indicate the occurrence of two-photon effects.

\begin{figure}
    \centering
    \includegraphics[width=0.33\textwidth]{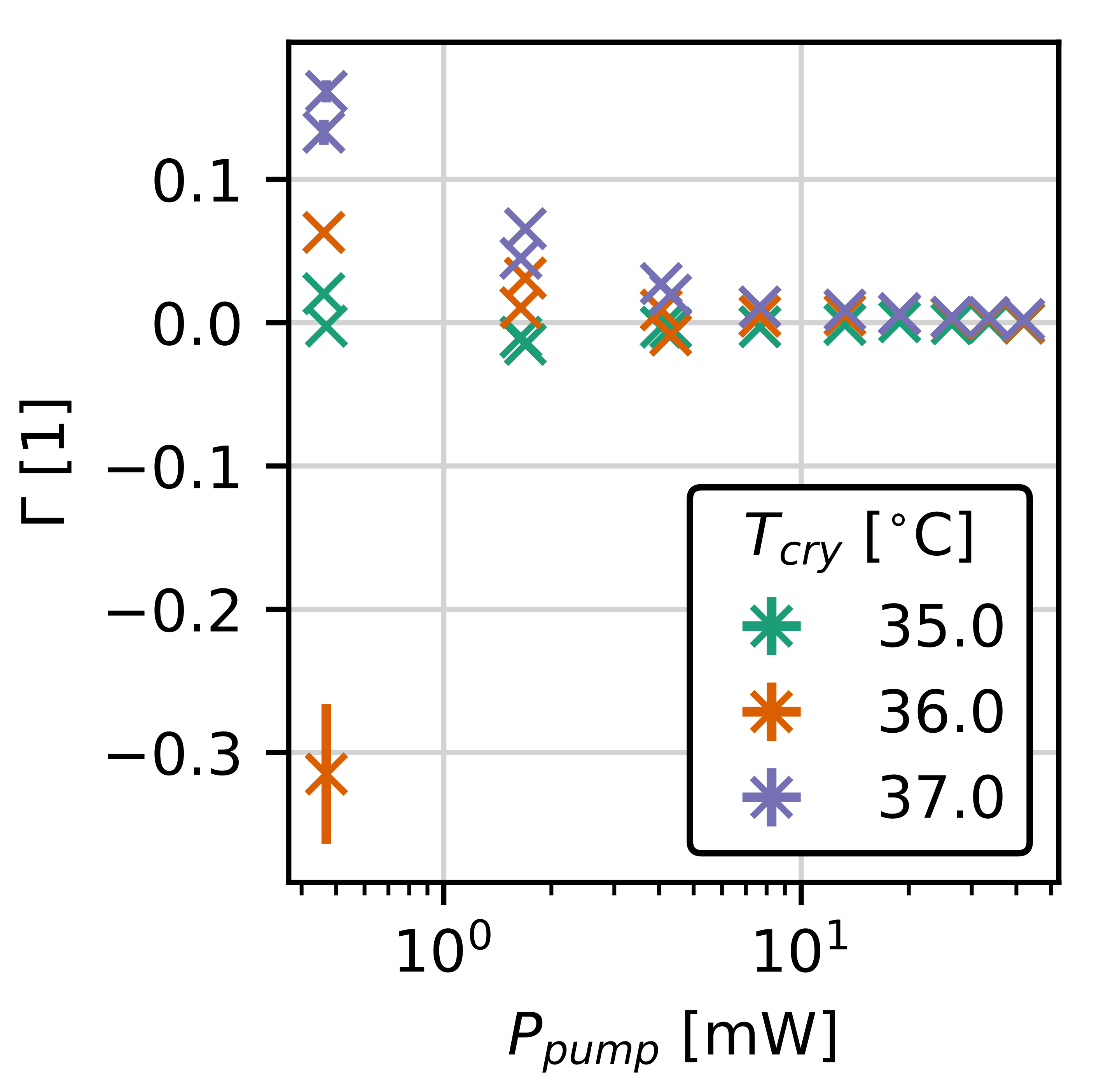}
    \caption{Biphoton ratio $\Gamma$ in dependency on the average pump power. Data points close to zero indicates that two-photon effects were not present according to the argumentation of Ref.{~\cite{Corona-Aquino2022}}.}
    \label{fig:Biphoton ratio}
\end{figure}

Fig.~{\ref{fig:Biphoton ratio}} shows the biphoton ratio $\Gamma$ related to the measurement data of Fig.~{\ref{fig:Absorption Rates}}(a). As visible, all data points are close to zero. Following the argumentation in Ref.{~\cite{Corona-Aquino2022}}, it can therefore be assumed that two-photon effects play a minor role for our measurements. Only at low power levels, $\Gamma$ slightly differs from zero and possibly indicate two-photon absorption. 

To sum up, it must be emphasized that the data base for absorption measurements in this study is not sufficient for a statement about general validity of the developed model. In particular, the visible qualitative behavior of the experimental data corresponds to the temperature of the maximum absorption only. For this reason, we think that our model represents a manageable starting point for predictions on eTPA even if further investigations are indispensable to provide clarification about the limits of this model.

\section{Conclusion}

In this work, we presented a straightforward and easily manageable theoretical approach to qualitatively predict the behavior of eTPA probabilities of the fluorescent dye Nile Red. By utilizing known information on the energy structure of the fluorophore in question, and supplementing it with reasonable assumptions for quantities not readily available in the literature, e.g., intermediate state widths and transition dipole moments, we were able to formulate a spectral response function of the fluorophore, against which the spectral properties of an incoming photon-pair state were optimized. We specifically focused on entangled photon pairs generated by a temperature-controlled SPDC source and found that the excitation probability of eTPA has a strong dependence on the phase matching conditions of the SPDC process and is influenced and optimized by tuning the temperature of the source which controls the phase matching. In particular, we showed that the maximum absorption probability for the investigated case of Nile Red and the particular photon-pair source is achieved in cases where the photon pairs are slightly nondegenerate.

Although we were using experimentally and numerically determined values given by former studies to establish the fluorophore's response, these can also be obtained by detailed quantum chemical calculations resulting in a complete ab-initio theory. In the latter case, the theory could then provide quantitative, as well as qualitative predictions. Also, the optimization procedure can be easily adapted to other sources of entangled photons, where the available parameters for influencing the photons' spectral properties would then be used to optimize the eTPA behavior.

We also supported our theoretical analysis by an experimental investigation of Nile Red in which we measured the dependence of the eTPA cross-section on the phase matching temperature of the SPDC crystal. The obtained results followed the general predictions of the model only partially. While the phase-matching temperature with the largest absorption cross-section agrees between experiment and theory, the changes in the magnitude of order differ strongly in contrast. However, the missing emission of fluorescence and the seeming absence of two-photon effects impeded a final proof of the occurrence of eTPA and, thus, the validity of our model. A deeper theoretical investigation to include parasitic effects and imaginable chemical influences, for example aggregation effects at high concentrations \cite{Lu1986}, as well as a broader experimental study are necessary to confirm the validity of our model.

\section*{Author Contributions}

Conceptualization, A.K. and T.B.G.; methodology, all authors; investigation, A.K. and T.B.G.; model development and calculations, A.K.; measurements, T.B.G., N.J. and P.T.; data curation, A.K. and T.B.G.; writing the manuscript, A.K. and T.B.G.; supervision, V.F.G., S.S., F.S, C.E. and M.G.; project administration, T.B.G., F.S., C.E. and M.G.; funding acquisition, F.S., C.E. and M.G.

A.K. and T.B.G. contributed equally to this work. All authors have read and agreed to the published version of the manuscript.

\section*{Funding}

This research was funded by Bundesministerium für Bildung und Forschung (BMBF) through the projects "LIVE2QMIC", funding ID 13N15954, and "QuantIm4Life", funding ID 13N14877.

The Leica Stellaris 8 (used for the fluorescence measurements) was funded by Deutsche Forschungsgemeinschaft (DFG, German Research Foundation), funding ID 460889961.

The superconducting nanowire single-photon detectors (used for the absorption measurements in the facility of PicoQuant GmbH) were funded by ATTRACT Phase 2 through the project "MicroQuaD-Material science", funding ID 101004462.

We further acknowledge funding for the Microverse Imaging Center by the DFG under Germany´s Excellence Strategy - EXC 2051 - project-ID 390713860 and the SFB Polytarget - project number 316213987 – SFB 1278.

Sina Saravi acknowledges funding by the Nexus program of the Carl-Zeiss-Stiftung (project MetaNN).

\section*{Data Availability Statement}

The data presented in this study are available upon request at the corresponding authors.

\begin{acknowledgement}

The authors thank the company PicoQuant GmbH, in particular Corinna Nock, Thomas Sch\"onau, and J\"urgen Breitlow, for providing a prototype of the used laser and opportunity to execute the experiments in their facility.

\end{acknowledgement}

\bibliography{References}

\end{document}